\documentclass[prb,aps,showpacs,showkeys,twocolumn,superscriptaddress,10pt,floatfix]{revtex4-1}

\usepackage{graphicx}
\usepackage{dcolumn}
\usepackage{bm}
\usepackage{color}
\usepackage{natbib}
\usepackage{overpic}
\usepackage{ulem}
\usepackage[colorlinks=true,linkcolor=blue]{hyperref}
\usepackage{verbatim}
\usepackage{amsmath}

\begin{document}

\title{Collective magnetic dynamics in artificial spin ice probed by AC susceptibility}

\author{Merlin Pohlit}
\email{merlin.pohlit@physics.uu.se}
\affiliation{Department of Physics and Astronomy, Uppsala University, Box 516, SE-75120 Uppsala, Sweden}

\author{Giuseppe Muscas}
\altaffiliation[Current address: ]{Department of Physics, University of Cagliari, S. P. Monserrato-Sestu km 0,700, 09042 Monserrato (CA), Italy.}
\affiliation{Department of Physics and Astronomy, Uppsala University, Box 516, SE-75120 Uppsala, Sweden}

\author{Ioan-Augustin Chioar}
\affiliation{Department of Physics and Astronomy, Uppsala University, Box 516, SE-75120 Uppsala, Sweden}

\author{Henry Stopfel}
\altaffiliation[Current address: ]{Department of Materials Science and Engineering, Uppsala University, Box 534, SE-751 21 Uppsala, Sweden.}
\affiliation{Department of Physics and Astronomy, Uppsala University, Box 516, SE-75120 Uppsala, Sweden}

\author{Agne Ciuciulkaite}
\affiliation{Department of Physics and Astronomy, Uppsala University, Box 516, SE-75120 Uppsala, Sweden}

\author{Erik \"Ostman}
\affiliation{Department of Physics and Astronomy, Uppsala University, Box 516, SE-75120 Uppsala, Sweden}

\author{Spyridon D. Pappas}
\altaffiliation[Current address: ]{Fachbereich Physik and Forschungszentrum OPTIMAS, Technische Universit\"at Kaiserslautern, 67663 Kaiserslautern, Germany.}
\affiliation{Department of Physics and Astronomy, Uppsala University, Box 516, SE-75120 Uppsala, Sweden}

\author{Aaron Stein}
\affiliation{Center of Functional Nanomaterials, Brookhaven National Laboratory, P.O. Box 5000, Upton, New York 11973, USA}

\author{Bj\"orgvin Hj\"orvarsson}
\affiliation{Department of Physics and Astronomy, Uppsala University, Box 516, SE-75120 Uppsala, Sweden}

\author{Petra E. J\"onsson}
\affiliation{Department of Physics and Astronomy, Uppsala University, Box 516, SE-75120 Uppsala, Sweden}

\author{Vassilios Kapaklis}
\email{vassilios.kapaklis@physics.uu.se}
\affiliation{Department of Physics and Astronomy, Uppsala University, Box 516, SE-75120 Uppsala, Sweden}

\date{\today}

\begin{abstract}
We report on the study of the thermal dynamics of square artificial spin ice, probed by means of temperature and frequency dependent AC susceptibility. 
Pronounced influence of the inter-island coupling strength was found on the frequency response of the samples. Through the subsequent analysis of the frequency- and coupling-dependent freezing temperatures, we discuss the phenomenological parameters obtained in the framework of Vogel-Fulcher-Tammann law in terms of the samples microscopic features. 
The high sensitivity and robust signal to noise ratio of AC susceptibility validates the latter as a promising and simple experimental technique for resolving the dynamics and temperature driven dynamics crossovers for the case of artificial spin ice.
\end{abstract}
\maketitle

\section{\label{sec:level1}Introduction}

Artificial Spin Ice (ASI), i.e. arrays of magnetostatically coupled ferromagnetic islands -- {\it mesospins} \cite{Ostman_1DIsing_2018} -- fabricated by nanolithography \cite{Wang2006,Nisoli2013,Heyderman:2013gb,Rougemaille_Colloquium_2019}, exhibit collective phenomena, and importantly, their interaction strength and geometry can be tailored almost at will \cite{ThermalFluctuationsInASI,Perrin_Nature_2016,Ostman_natphys_2018,Nisoli_NatPhys_Perspective, Pohlit_JAP_2015, Pohlit_JAP_2016}. Properly designed to support thermal fluctuations, ASI systems can serve as a platform for the investigations of thermal magnetization dynamics and freezing transitions in tailored nanostructures, which can also be used to mimic the dynamical properties of frustrated, naturally-occurring magnetic spin systems \cite{melting_artificial_spin_ice,Morgan_natphys_2010,Farhan:2013ki,ThermalFluctuationsInASI,Porro:2013cm,Ostman_natphys_2018}. Insights into the freezing transition and the nature of the frozen low temperature states were obtained by investigations using magnetometry \cite{Andersson2016} and synchrotron-based scattering- and microscopy-techniques\textcolor{green}{ \cite{Morley2017, Sendetskyi2019, Saccone2019}}. With the exception of early work based on temperature dependent magneto-optical measurements  \cite{melting_artificial_spin_ice} and more recent works using synchrotron-based magnetic microscopy  \cite{ThermalFluctuationsInASI} and muon relaxation  \cite{LJHeyderman,Leo:2018di}, experimental studies of thermally induced transitions are scarce. Furthermore, experimental techniques based on synchrotron radiation and muons impose limitations on availability and accessible time-scales. To this end, AC susceptibility is a well established and accessible technique for probing magnetization dynamics, giving access to a wide frequency range \cite{Bedanta2008, Topping_2019_review}.

In this work, we report on AC susceptibility measurements of thermally active extended square ASI arrays measured in a wide frequency and temperature range. We study arrays that are composed of close to identical mesospins, with different gaps between the elements, in order to explore the influence of coupling strength on the collective dynamics. Exploring the frequency dependence of the AC susceptibility signal we employ the VFT law, that can be used for describing the low-field magnetic relaxation of weakly interacting nanoparticle systems \cite{LandiJAP2013,Vernay_PRB_2014} but has recently been applied also to ASI systems\cite{Andersson2016, Morley2017}, attempting to extract parameters that can be directly related to the magnetostatic energies of the ASI arrays. We discuss the validity of this simplified approach and address the limitations of such models, in the framework of thermal ASI arrays.

\section{Experimental Details}

\begin{figure}[ht]
	\includegraphics[width=1\columnwidth]{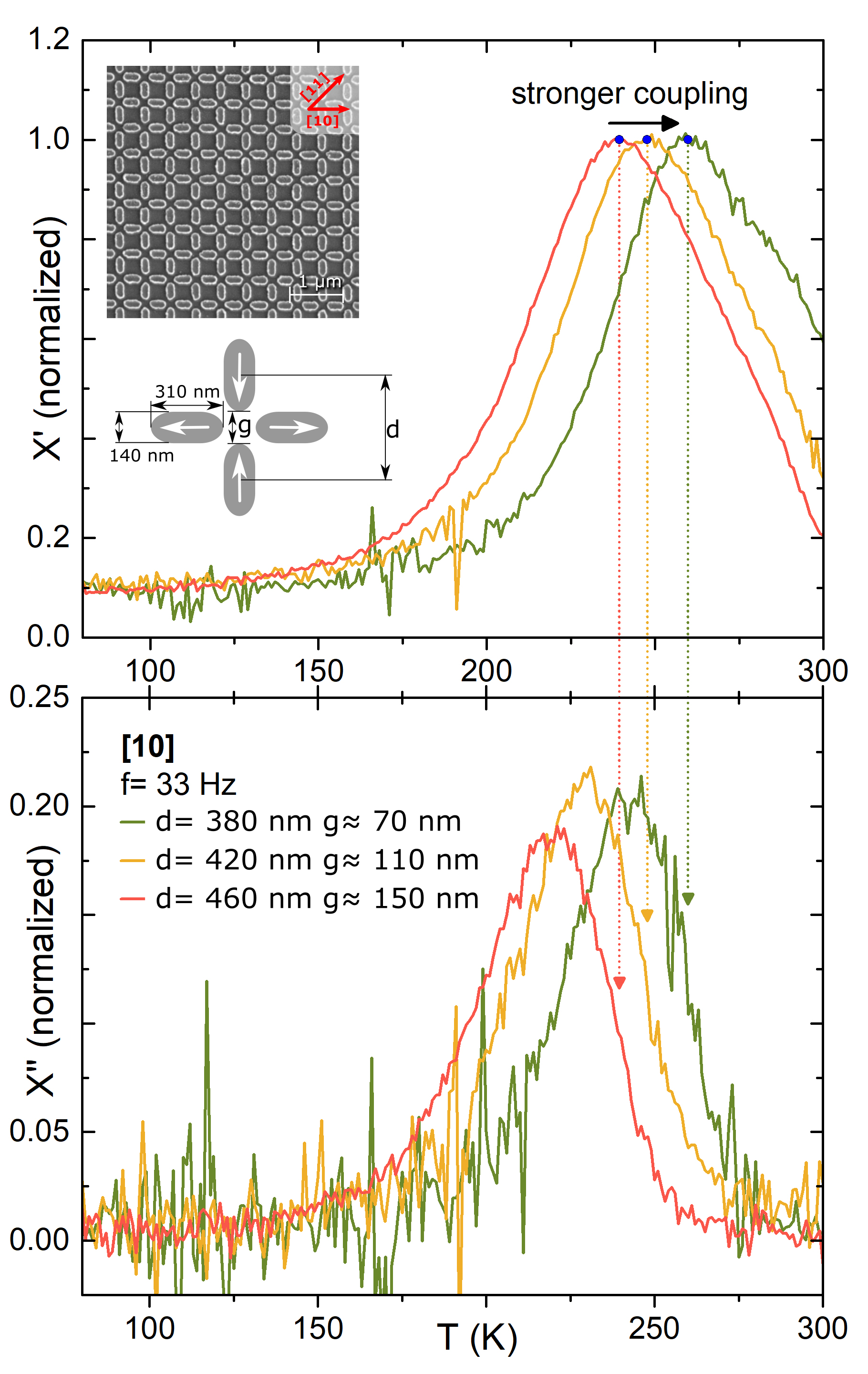} 
	\caption{Temperature dependence of the in-phase component of the AC susceptibility $\chi'$ (upper panel) and  out-of-phase component $\chi''$ (lower panel) for three  square ASI arrays with different inter-island coupling strengths.  The inset depicts the geometry and a typical SEM image of the nanostructures composed of 
	$310\pm 15\,\rm{nm} \times 140\pm 15\,\rm{nm}$ magnetic islands. Only the gap $g$, differs between the three investigated arrays, as the center-to-center distance $d$ is varied to be $d \approx 380\,\rm{nm}$, $420\,\rm{nm}$, or $460\,\rm{nm}$ respectively.
	The measurements were conducted with an AC magnetic field strength of $\mu_{\rm{0}}H_{\rm{ac}}=0.1\,\rm{mT}$ and a frequency of $f=33\,\rm{Hz}$ aligned along the [10]-direction of the array. For direct comparison the $\chi'$ and $\chi''$ curves were normalized to the value of the corresponding maximum of $\chi'$  ($T_{\rm{m}}$, blue dots).
	}
	\label{fig:fig1}
\end{figure}

The extended square ASI structures 
were produced by post-patterning of $\delta$-doped Pd(Fe) thin films \cite{parnaste2007dimensionality}, employing electron-beam lithography (EBL) followed by argon-ion milling. The films, consisting of $40\,\rm{nm}$ Palladium, 2.2 monolayers of Iron, and a $2\,\rm{nm}$ Palladium capping layer, were all grown on a Vanadium seeding layer on top of Magnesium oxide (MgO(001)) substrates by DC magnetron sputtering. 
The total effective thickness of the magnetic layer (Fe and magnetically polarized Pd) was previously estimated to be $2\,\rm{nm}$  \cite{parnaste2007dimensionality,Hase_PRB_Delta_polarizability_2014}.
Vibrating Sample Magnetometetry (VSM) revealed that the temperature dependence of the in-field volume magnetization can be described by: $M_{\rm{s}}(T) = M_{\rm{0}}(1-T/T_{\rm{0}})^{0.5}$, with a Curie temperature $T_{\rm{0}} = 410\,\rm{K}$ \cite{Andersson2016}. The size of the islands, the lattice parameter and the distance between the islands were determined after the EBL process using Scanning Electron Microscopy (SEM). A typical SEM image is shown as an inset of Fig.~\ref{fig:fig1}. All the islands have a length of $310\pm15\,\rm{nm}$ and a width of $140\pm15\,\rm{nm}$. Arrays with different inter-island lattice spacing, $d$, of 380, 420 and 460 nm were prepared which yields different gaps between the islands ($g$ = 70, 110, and 150 nm). The difference in distances between the islands results in different magnetostatic coupling strengths of the mesopins.

The magnetic moment of a mesospin at $T = 5\,\rm{K}$ was determined to be $m_{\rm{0}} = M_{\rm{0}}V = 6.5 \times10^{6}\mu_{\rm{B}}$, where V is the volume of the magnetic material in the island.\footnote{The small difference between the moment stated here compared to \citet{Andersson2016} stems from a more detailed analyses of the island dimensions from SEM imaging.} Micromagnetic calculations using $\rm{MuMax}^3$ \cite{Vansteenkiste2014} revealed non-collinearities of the moment within the islands. A reduced effective moment  of approximately $m_{\rm{0,eff}} = 0.65 \times m_{\rm{0}} = 4.2 \times10^{6}\mu_{\rm{B}}$ at $T = 0\,\rm{K}$, is used to compensate for the dynamic non-collinear internal magnetic structure of the elements (see  \citet{Andersson2016}, \citet{Bessarab2012},  \citet{Bessarab2013} and \citet{Gliga_PRB_2015}).
Furthermore, the intrinsic effective moment $m_{\rm eff}(T)$ of the mesospins is temperature dependent\cite{ThermalFluctuationsInASI,Andersson2016,Ostman_natphys_2018}:

\begin{equation}
m_{\rm eff}(T)=m_{\rm 0,eff}(1-T/T_0)^{0.5}
\label{Moment_scaling}
\end{equation}

\noindent

The dynamic response of the three extended square ASI arrays (2$\times$2 mm$^2$ each) was investigated by AC susceptibility employing a magneto-optical Kerr effect (MOKE) magnetometer in longitudinal mode\cite{ASPELMEIER_AC_Susceptibility_1995}. For this purpose the samples were mounted in a cryostat with optical access ($4\,\rm{K} \leq T \leq 300\,\rm{K}$). All experiments were performed with a 20 mW laser with a wavelength of 660\,\rm{nm}. A pair of Helmholtz coils was used to generate a small sinusoidal magnetic field with a given frequency, aligned along the [10]-direction of the nanostructured array (see inset of Fig. \ref{fig:fig1}, upper panel). For frequencies between $f= 0.1-333\,\rm{Hz}$ an amplitude of $\mu_{\rm{0}}H_{\rm{ac}}=0.1\,\rm{mT}$ was used, while for the higher frequencies $f= 1111\,\rm{Hz}$ and $f= 3333\,\rm{Hz}$ the amplitude was reduced to $\mu_{\rm{0}}H_{\rm{ac}}=0.01\,\rm{mT}$. A lock-in amplifier (Stanford Research SR830) was used as a voltage source for generating the magnetic field and measuring the AC susceptibility at the corresponding frequency. The sample was shielded from the earth's magnetic field by a double-wall mu-metal cylinder, and demagnetization 
was performed prior to each cool-down. 

Applying the magnetic field along the [10]-direction of the samples (see Fig.  \ref{fig:fig1} ) in the longitudinal MOKE configuration probes primarily the mesospins with long axes parallel to the direction of the magnetic field. Both in- and out-of-phase components, $\chi'$ and $\chi''$, were recorded during warming  up to the maximum temperature of $T= 300\,\rm{K}$, starting either at $T\approx 80\,\rm{K}$ ($f=33\,\rm{Hz}$) or $T=200\,\rm{K}$ (other frequencies) and by stabilizing the cryostat in discrete $1\,\rm{K}$ temperature steps while keeping the amplitude and the frequency of the magnetic field fixed. In every step, sufficient time (60 s) was taken to allow the measurement system (cryostat) to stabilise before starting a measurement \footnote{As temperature gradients between sample surface and temperature sensor of the set-up cannot be fully excluded it is plausible to assume the presence of temperature drifts of the sample of up to $1\,\rm{K}$.}. The time for data acquisition at each temperature step was increased for lower frequencies to ensure an acceptable signal to noise ratio. In this respect it should be noted that the magnetic signal of the samples is extremely small, as it originates from a 2.2 ML magnetic Fe layer in Pd, with a coverage of about $\approx 37-54\%$ (depending on the geometry of the pattern). For the weakest coupled sample ($d = 460\,\rm{nm}$) this equals an effective magnetic coverage of a sub-monolayer thick continuous Fe film in Pd. 

\section{Results and discussion}

\begin{figure}[t]
	\includegraphics[width=\columnwidth]{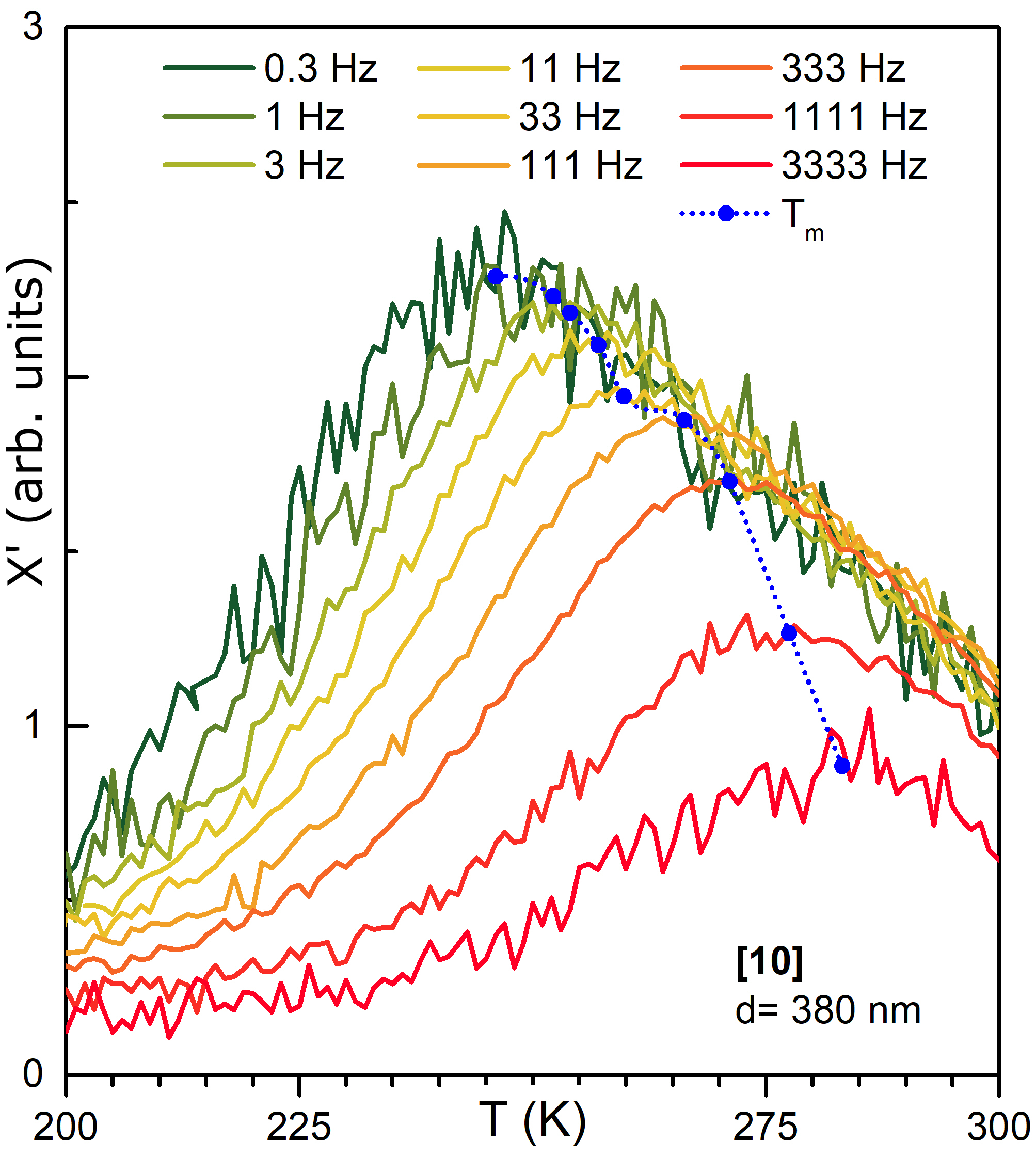}%
	\caption{\label{fig:Magneto_Fits} Temperature dependence of the in-phase component $\chi'$ of the AC susceptibility measured at different excitation field frequencies. The magnetic susceptibility was recorded while heating the sample. Typical dataset obtained on the array with the strongest inter island-interactions ($d=380\,\rm{nm}$) using excitation fields of $\mu_{\rm{0}}H_{\rm{ac}}=0.1\,\rm{mT}$ ($f=0.1-333\,\rm{Hz}$) and $\mu_{\rm{0}}H_{\rm{ac}}=0.01\,\rm{mT}$ ($f=1111-3333\,\rm{Hz}$) oriented along [10]-direction. The curves are normalized to their respective magnetic field amplitudes. The blue dots mark the position of the maximum $T_{\rm{m}}$. The dotted blue line is a guide to the eye.}
	\label{fig:fig2}
\end{figure}

For each of the three studied arrays, a single peak in both $\chi'$ and $\chi''$ is observed, as illustrated in Fig.  \ref{fig:fig1}. The shape of the peaks is similar for the three arrays, while a shift in the peak position $T_{\rm{m}}$ towards higher temperatures is observed with decreasing inter-island distance, $i.e.$ increasing inter-island coupling strength. We attribute the maximum in the AC susceptibility to the condition when the average relaxation time is equal to the observation time window $\tau_{\rm{m}}(T_{\rm{m}}) = 1/(2\pi f)$\cite{souletie_critical_1985}, where $f$ is the frequency of the applied magnetic field and $T_{\rm{m}}$ is the corresponding temperature commonly referred to as the blocking temperature in superparamagnetic samples. Consequently, by determining the peak positions $T_{\rm{m}}$ for different observation time windows (frequencies), the average relaxation time of the system as a function of temperature can be extracted from the AC susceptibility data. 

In order to investigate both the temperature dependence of the relaxation time and the effect of inter-island interactions, $T_{\rm{m}}$ was determined for different frequencies, effectively resulting in different observation time windows. 
Typical results are illustrated for the array with the strongest interactions ($d = 380\,\rm{nm}$) in Fig.  \ref{fig:fig2}. $T_{\rm{m}}$ was determined by fitting a parabola to a region of interest around the maximum of each curve. For each of the three arrays, nine $T_{\rm{m}}$ values are extracted, corresponding to the employed excitation frequencies $f$, as illustrated in Fig. ~\ref{fig:fig2}. The peaks are found to shift towards higher temperatures with increasing frequency.

\begin{figure}[b]
	\includegraphics[width=\columnwidth]{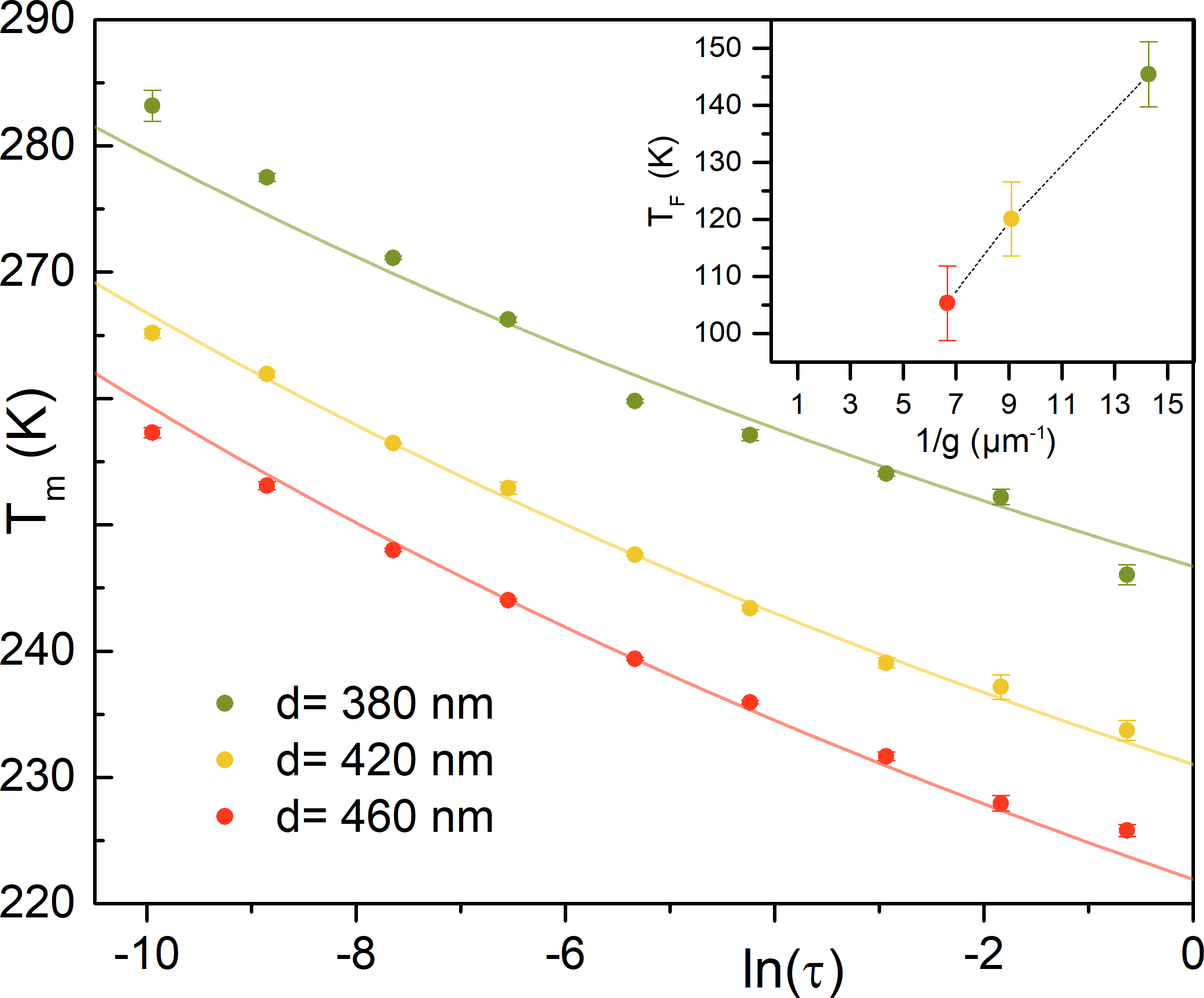} 
	\caption{Compiled $T_{\rm{m}}$ datasets, as obtained from the in-phase component $\chi'$ at nine frequencies $f$ for the three arrays, plotted against $ln(\tau_{\rm{m}}) =ln(1/(2\pi f))$. The bold lines represent fits based on the VFT law with a fixed $\tau_{\rm{0}}= 1\cdot10^{-11}\,\rm{s}$. The results of the fits are summarized in Table \ref{tab:table1}. Inset: Systematic reduction of the extracted freezing Temperatures $T_{\rm{F}}$ with increased island spacing i.e. reduced coupling strength.}
	\label{fig:fig3}
\end{figure}

\subsection{Fitting the experimental data using the Vogel-Fulcher-Tammann law}
We start the analysis by employing the empirical VFT law for the relaxation time $\tau$ of weakly interacting magnetic particles
\cite{Shtrikman1981}, an approach which has previously been  used to describe the relaxation in artificial spin ice structures  \cite{Andersson2016, Morley2017}. Within this approach the relaxation time can be calculated by

\begin{equation}
\tau = \tau_{\rm{0}} \cdot \exp{{\Big( \frac{E_{\rm{K}}}{k_{\rm{B}} (T-T_{\rm{F}})} \Big),}}
\label{eq:fulcher}
\end{equation}

\noindent
where $\tau_0$, $E_{\rm{K}}$, $k_{\rm{B}}$, $T$ and $T_{\rm{F}}$ correspond the inverse attempt frequency, the intrinsic energy barrier, the Boltzmann constant, the temperature and the freezing or Fulcher temperature, respectively. While the mesospin's energy barrier $E_{\rm{K}}$ depends on the shape and magnetisation of the mesospins, the Fulcher temperature $T_{\rm{F}}$ is indicative of the interaction strength of the elements. 
The intrinsic energy barrier $E_{\rm{K}}$ in Eq. (\ref{Moment_scaling}) is attributed to the shape anisotropy and its temperature dependence can be captured by:

\begin{equation}
   E_{\rm{K}} = \mu_{\rm{0}} \Delta N \frac{[m_{\rm{eff}}(T)]^2 V}{2} = \mu_{\rm{0}} \Delta N \frac{m_{\rm{0,eff}}^2 V}{2} (1-\frac{T}{T_0}),
   \label{Energy-barrier}
\end{equation}

\noindent 
where $\mu_{\rm{0}}$ is the vacuum magnetic permeability and $\Delta N$ is the island differential demagnetizing factor according to \citet{PhysRev.67.351}. Taking into account the temperature dependence of the island saturation magnetization $M_{\rm{S}}$ in the VFT law, the temperature dependence of the energy barrier becomes $E_{\rm{K}} = E_{\rm{K}}^{\rm{0}}\cdot(1-T/T_{\rm{0}})$, where $E_{\rm{K}}^{\rm{0}}=\mu_{\rm{0}} \Delta N \frac{m_{\rm{0,eff}}^2 V}{2}$ is the energy barrier at zero temperature.


Fig.  \ref{fig:fig3} presents the extracted $T_{\rm{m}}$ values along with their corresponding VFT fits. The fitting was performed by reversing Eq. (\ref{eq:fulcher}), i.e. solving for $T$, and using a Levenberg-Marquardt algorithm to find the best fit for $E_{\rm{K}}^{\rm{0}}$ and $T_{\rm{F}}$. This procedure facilitates weighting the data points with their respective uncertainties on the maxima positions $T_{\rm{m}}$, while the uncertainty in the magnetic field frequency given by the lock-in amplifier was considered negligible.\footnote{Note that, both for the parabola as well as the nonlinear VFT-fits, the weighting is defined by the inverse quadratic form of the individual errors for each data point, i.e. $w_{\rm{i}}=1/\sigma_{\rm{i}}^{2}$. Furthermore, the standard errors of the fitting parameters are scaled with the square root of the reduced chi squared.}
The uncertainties of $T_{\rm{m}}$ were taken from the fits used for determining the maximum positions, while the uncertainty in selecting the individual temperature ranges used for peak finding were considered negligible. This approach qualitatively captures the effect that the uncertainty is larger for the highest and lowest frequencies measured. 
Since all arrays are composed of the same size elements, the flipping time was assumed to be constant, $\tau_{\rm{0}}=10^{-11}\,\rm{s}$, in accordance with a previous relaxation study \cite{Andersson2016}. The same energy barrier, $E_{\rm{K}}^{\rm{0}}$ was used for all three gap sizes studied. 
The summary of the results from these analysis is found in Table \ref{tab:table1}.

For comparison, an independent magnetostatic estimation of the intrinsic energy barrier was made, calculating $\Delta N$ using the Osborn methodology \cite{PhysRev.67.351}. The effective magnetic moment of an island $m_{\rm{0,eff}}$ was used in order to take the non-collinearities of the moment within the islands into account. By using this approach and the temperature dependence of $m_{\rm{eff}}$, the mean value of the energy barrier was determined to be $E_{\rm{K}}/k_{\rm{B}} \approx 3200\,\rm{K}$ at $250\,\rm{K}$. Using the temperature-scaling for the fitted value of $E_{\rm{K}}^{\rm{0}}$, we obtain $E_{\rm{K}}/k_{\rm{B}}(T= 250\,\rm{K}) = 2510\,\rm{K} \pm 120\,\rm{K}$, a value which, for the current case, is not so far from the magnetostatic estimation.

\subsection{On the extraction of characteristic interaction energies}

With a single energy barrier fitted for all three datasets, the impact of the different gap sizes is inherently accounted for by the Fulcher temperature, $T_{\rm{F}}$. This dependence is represented in the inset of Fig.  \ref{fig:fig3}, highlighting a connection between $T_{\rm{F}}$ and the inter-island magnetostatic interactions. This further raises the question of whether the Fulcher temperature can be systematically and accurately used to extract the characteristic interaction strength between mesospins.

In the framework of a weak-coupling regime in spin glasses, \citet{Shtrikman1981} offered a recipe for extracting the typical interaction energy between the magnetic components by using the mean-field based formula:

\begin{equation}
   E_{\rm{i}}^0 = \sqrt{k_{\rm{B}} \cdot T_{\rm{F}} \cdot E_{\rm{K}}^0} \\
   \label{eq2}
\end{equation}
where $E_{\rm{K}}^0$ represents the average intrinsic energy barrier of a single magnetic element, while $E_{\rm{i}}^0$ is the characteristic interaction energy between the magnetic constituents, in turn defined as $E_{\rm{i}}^0 = \mu_{\rm{0}} m_{\rm 0,eff} \cdot H_{\rm{i}}^0$, with $H_{\rm{i}}^0$ representing the characteristic mean interaction field. Note that the latter quantity is determined by considering the ground state manifold of the magnetic system, i.e. the configurations for which the mean interaction field is the strongest. 

\begin{figure}
	\includegraphics[width=\columnwidth]{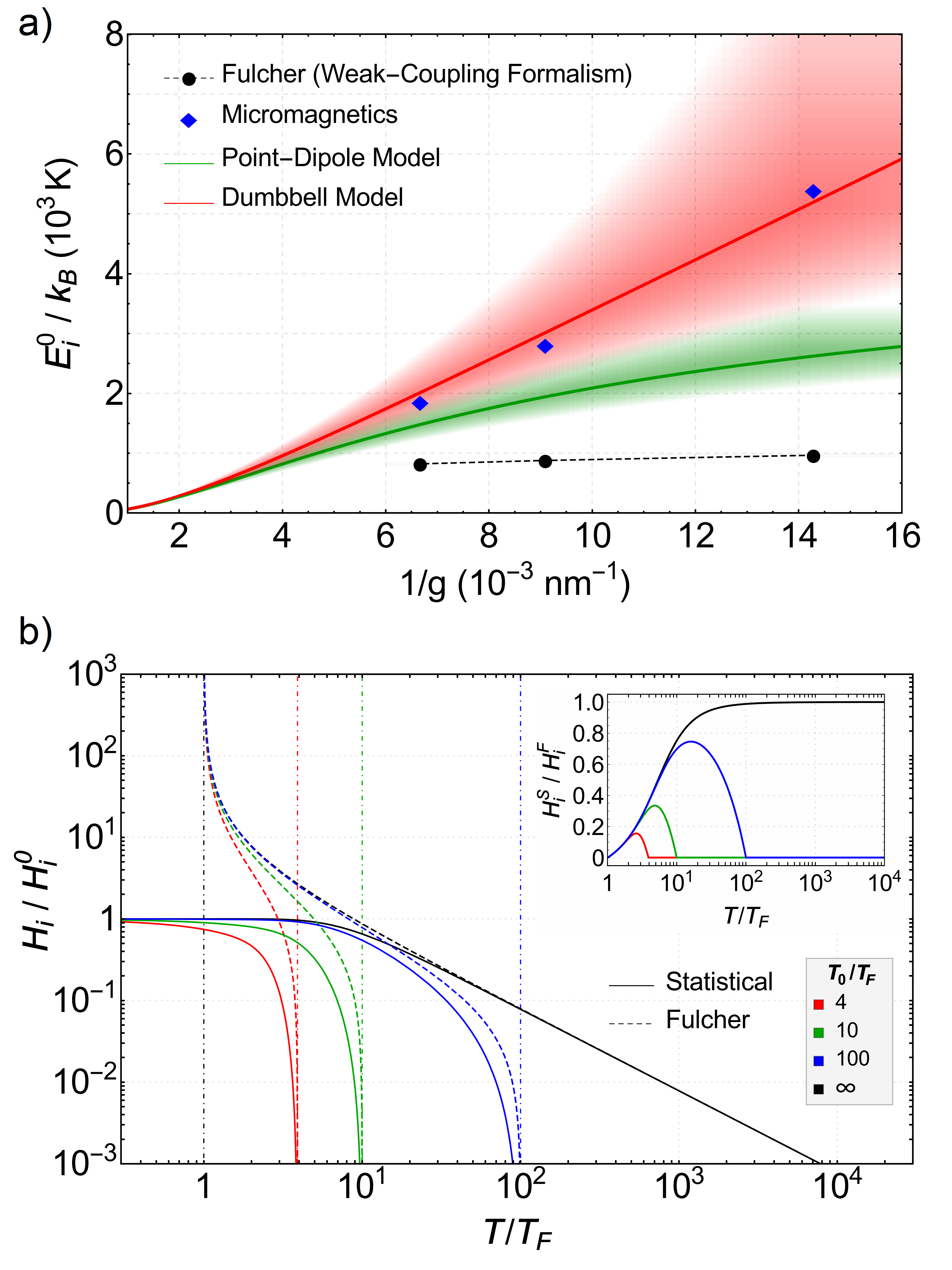} 
	\caption{(a) Comparison between the interaction energies per island ($E_i^0$) extracted for the three experimental gap sizes (black dots) and the type I tilling ground-state interaction energies per island computed for various inverse gap sizes ($1/g$) using micromagnetics (blue diamonds), the dumbbell model (red line) and the point-dipole model (green line) for the given sample geometries. The shaded areas account for the error originating from a gap uncertainty of $\pm 15\,\rm{nm}$, extracted from the analysis of the SEM images. (b) Normalized temperature dependence of the normalized mean interaction field for various thermal dynamics ranges, calculated using the Fulcher-based expression (dotted lines) and the statistical approach (continuous lines). The black lines correspond to the case with temperature-independent energy barriers and island moments, while the coloured lines consider the temperature scaling for different normalized Curie temperatures $T_0/T_F$. The vertical dashed lines mark the boundaries of each considered range. The inset gives the normalized temperature dependence of the ratio between the Fulcher-based and statistical-based interaction fields. The experimental values provided in Table \ref{tab:table1} for $E_{\rm{K}}^{\rm{0}}$ and $T_{\rm{F}}$ are used in both figures.}
	\label{fig:fig4}
\end{figure}

We have extracted the characteristic interaction energies for the three studied arrays, presented in Table~\ref{tab:table1}, and compared them with the corresponding ground-state energies of a square spin ice, i.e. type I tiling, given by conventional interaction models based on the point-dipole approximation, the dumbbell representation, as well as micromagnetic simulations \footnote{The micromagnetic simulations have been performed using the $\rm{MuMax}^3$ GPU-accelerated micromagnetic simulation program \cite{Vansteenkiste2014}. The simulations were performed using the following parameters: $M=499289\,\rm{A/m}$, $A_{\rm{ex}}=3.36\cdot10^{-12} \rm{J/m}$, length $=310\,\rm{nm}$, width $=140\,\rm{nm}$, and a thickness of $2\,\rm{nm}$.}. As shown in Fig.~\ref{fig:fig4}, the VFT-law obtained values deviate considerably from the corresponding estimations of all the models employed, with resulting energies that can be more than 5 times smaller than expected from micromagnetic estimates. Furthermore, although limited by the number of experimentally-available points, the gap-dependent scaling of the extracted values appears to be much less pronounced than the model predictions. This approach was previously also employed by \citet{Morley2017} - attempting to extract experimental values of the characteristic mesospin interaction energies - who reported a significant underestimation of the extracted energies.

\begin{table}[t]
	\caption{\label{tab:table1}%
		Left: Energy barrier $E_{\rm{K}}^{\rm{0}}$ and Fulcher temperature $T_{\rm{F}}$ as extracted from the VFT fits of the experimental data plotted in Fig.  \ref{fig:fig3}, assuming a fixed $\tau_{\rm{0}}=1\cdot10^{-11}\,\rm{s}$, as done in \citet{Andersson2016} for the same samples. Right column: Characteristic magnetostatic interaction energies obtained from the VFT fits in the weak coupling formalism using Eq. (\ref{eq2}). 
		The given uncertainties correspond to the fitted parameter's standard errors as provided from the weighted non-linear fitting procedure described in the text.}
		\centering
	\begin{ruledtabular}
		\begin{tabular}{ccccc|cc}
			\textrm{$\rm{d}$}&\textrm{$\rm{gap}$}&
			\textrm{$E_{\rm{K}}^{\rm{0}}/k_{\rm{B}}$}&
			\textrm{$T_{\rm{F}}$}&&
			\textrm{$E_{\rm{i}}^{\rm{0}}/k_{\rm{B}}$}\\
		    \textrm{(nm)}&
	    	\textrm{(nm)}&
			\textrm{(K)}&
			\textrm{(K)}&&
			\textrm{(K)}\\
			\colrule
			380 & $70$ &                 & $145 \pm 6$ && $966 \pm 30$ \\
			420 & $110$ & $6438 \pm 308$ & $120 \pm 6$ && $879 \pm 32$ \\
			460 & $150$ &                & $105 \pm 7$ && $822 \pm 32$ \\
		\end{tabular}
	\end{ruledtabular}
\end{table}

While several factors can contribute to the mismatch between the experimental values and the various models, we mainly attribute these discrepancies to the incompatibility between the frameworks for Eq. (\ref{eq2}) and artificial spin systems. 
The basic assumption of the formalism is that the VFT law can be treated as an interacting Arrhenius law, i.e. an Arrhenius-like expression in which the intrinsic energy barrier is further biased by a local interaction field: 

\begin{equation}
\ln{\Big (\frac{\tau}{\tau_{\rm{0}}} \Big )} = \frac{E_{\rm{K}}}{k_{\rm{B}} (T-T_{\rm{F}})} = \frac{E_{\rm{K}}+\mu_{\rm{0}} m_{\rm 0,eff} \cdot H^{\rm{F}}_{\rm{i}}(T)}{k_{\rm{B}} T}
\label{eq3}
\end{equation}
where $H^{\rm{F}}_{\rm{i}}(T)$ represents the temperature-dependent mean interaction field and $E_{\rm{K}}$ is generally assumed to be a temperature-independent quantity. This equality imposes a certain temperature dependence on this variable, parameterized by $E_{\rm{K}}$ and $T_{\rm{F}}$:
\begin{equation}
H^{\rm{F}}_{\rm{i}}(T) = \frac{1}{\mu_{\rm{0}} m_{\rm 0,eff}} \cdot \frac{E_{\rm{K}} \cdot T_{\rm{F}}}{T-T_{\rm{F}}}
\label{eq4}
\end{equation}
Notice that this interaction field presents a divergence around the Fulcher temperature, a physically unrealistic feature. Taking now an at-equilibrium statistical perspective, the mean interaction field should converge at low temperatures towards a finite value, $H^{\rm{0}}_{\rm{i}}$, corresponding to the ground-state manifold. We shall further consider the choice of \citet{Shtrikman1981} assuming a hyperbolic tangent behavior, akin to the case of a paramagnetic system in an externally applied field:

\begin{equation}
H^{\rm{S}}_{\rm{i}}(T) = H^{\rm{0}}_{\rm{i}} \cdot \tanh{\Big ( \frac{\mu_{\rm{0}} m_{\rm 0,eff} \cdot H^{\rm{0}}_{\rm{i}}}{k_{\rm{B}} T} \Big )}
\label{eq5}
\end{equation}
This brings us to the remaining conditions for the validity of Eq. (\ref{eq2}): 
\begin{itemize}
   \item $T \gg T_{\rm{F}}$,  i.e. the temperature should be sufficiently far away from the Fulcher temperature.
   \item  $k_{\rm{B}} T \gg E^{\rm{0}}_{\rm{i}}$,  i.e. the energy  associated to the thermal bath should be much larger than the characteristic interaction energy.
\end{itemize} 
It should also be noted that the intrinsic energy barrier is also assumed to be much higher than both the thermal bath and the interaction energy, i.e. $E_K \gg k_{\rm{B}} T$ and $E_K \gg E^0_i$. With these in place, the functions describing the two interaction fields, $H^{\rm{F}}_{\rm{i}}(T)$ and $H^{\rm{S}}_{\rm{i}}(T)$, become compatible to a first order expansion. This validity region is illustrated in Fig. \ref{fig:fig4}(b). Notice how the two interaction fields present almost identical temperature scaling once the temperature is about two orders of magnitude higher than $T_{\rm{F}}$.

If we now consider the temperature dependence of the intrinsic energy barrier and the magnetic moment, a key feature of mesoscopic spin systems, the possibility of finding a compatibility region for the two interaction fields is severely limited.
Fig. \ref{fig:fig4}(b) illustrates the impact of having a range in thermal dynamics bound by the Fulcher and Curie temperatures. Here, the energy barrier, $E_{\rm{K}}$, from Eq. (\ref{eq3}) is replaced with its temperature dependent form given by Eq. (\ref{Energy-barrier}), while the characteristic interaction field from Eq. (\ref{eq5}), $H_{\rm{i}}^0$, is similarly replaced with a temperature dependent expression, $H_{\rm{i}}^0 \cdot (1-T/T_0)^{0.5}$, thus accounting for the scaling of the inter-island couplings. As it can be seen, only for a thermal range spanning several orders of magnitude between $T_0$ and $T_{\rm{F}}$, can one achieve a compatibility region that accommodates the aforementioned formalism. This is particularly highlighted by the inset of Fig. \ref{fig:fig4}(b), where the ratio between the two interaction field expressions is plotted as a function of temperature. The red lines correspond to the most weakly interacting sample, with an average gap of 150 nm and characterized by a $T_0/T_{\rm{F}} \cong 4$ ratio. Even for this scenario, there is no clear overlap between the two mean interaction fields within the experimental temperature window, which should therefore compromise the matching with the interaction models considered in Fig. \ref{fig:fig4}(a).

A microscopic description of the phenomenological parameter $T_{\rm{F}}$ was provided in a study of the AC susceptibility of weakly interacting magnetic nanoparticles by  \citet{Vernay_PRB_2014}. 
The VFT law, assuming the case of both weak dipolar interactions and surface anisotropy for the magnetic nanoparticles, can be transformed into a semi-analytical expression, linking $T_{\rm{F}}$ to deviations from uniaxial anisotropy and dipole-dipole interactions\cite{Vernay_PRB_2014}.
While this analysis assumes weak interactions being valid also in our case, the modelling of the interactions with discrete point dipoles is an imprecise description of our spatially extensive thermal mesospins (see Fig. \ref{fig:fig4}(b)). On the other hand this approach\cite{Vernay_PRB_2014} can serve as a stimulation for the development of a new and revised formalism capable of capturing in detail the collective temporal\cite{Andersson2016,Ostman_natphys_2018} and thermal dynamics\cite{Pohlit_JAP_2015,Pohlit_JAP_2016,ThermalFluctuationsInASI,Ostman_natphys_2018} of artificial spin ice.

\section{Conclusions}
We studied the AC susceptibility of thermally active square ASI arrays of varying interaction strength. The freezing of the mesospin dynamics was measured over a wide range of observation times and a systematic dependence of the freezing temperature on the inter-island coupling strength was found. Extracting magnetostatic energies from the frequency dependence using the VFT law, which was recently applied to ASI\cite{Andersson2016, Morley2017}, revealed significant discrepancies of the obtained interaction energies compared to theoretical estimates, similar to the case of \citet{Morley2017}. Besides experimental uncertainties, we attribute this to the violation of the requirements of weak coupling to extract energies in the VFT formalism, along with the inability to obtain measurements at temperatures far above the freezing transition, while staying well below the material's Curie temperature. We note that these requirements are generally difficult to meet in thermally active mesospin systems. Therefore a more advanced model enabling the extraction of microscopic variables and accounting for the details of our mesospin systems, such as the internal magnetic structure, temperature dependence of the energy barriers and interaction energies, is highly desirable and will be developed in future works.

Nonetheless, AC susceptibility using a longitudinal MOKE setup was proven to be a simple yet powerful technique for studying magnetization dynamics of thermally active nanostructures. Its high sensitivity allows investigations of minute changes in the mesospin dynamics in arrays with an intrinsically low magnetic moment. The method lends itself to temperature and frequency dependent studies on a laboratory scale\cite{Topping_2019_review}, which facilitates an evaluation of well-established models, or scaling laws for the description of mesospin systems. The method can further serve as an excellent tool for the characterization of thermal dynamics, collective behaviour \cite{Gliga_PRL_2013,PhysRevB_Agne_FMR_2019} and thermodynamic phase transitions in magnetic metamaterials \cite{Heyderman:2013gb,Leo:2018di}, such as ASIs \cite{LJHeyderman,Ostman_natphys_2018}.
\newline

The data that support this study are available via the Zenodo repository \cite{zenodo_susceptibility}.

\begin{acknowledgments}
The authors thank Prof. Per Nordblad (Department of Engineering Sciences, Uppsala University) for fruitful and stimulating discussions. The authors acknowledge support from the Swedish Research Council, the Swedish Foundation for International Cooperation in Research and Higher Education (STINT) project KO2016-6889 and the Knut and Alice Wallenberg Foundation project ``{\it Harnessing light and spins through plasmons at the nanoscale}'' (2015.0060). This research used resources of the Center for Functional Nanomaterials, which is a U.S. DOE Office of Science Facility, at Brookhaven National Laboratory under Contract No. DE-SC0012704.

\end{acknowledgments}

\bibliographystyle{apsrev4-1}

\begin{thebibliography}{42}%
\makeatletter
\providecommand \@ifxundefined [1]{%
 \@ifx{#1\undefined}
}%
\providecommand \@ifnum [1]{%
 \ifnum #1\expandafter \@firstoftwo
 \else \expandafter \@secondoftwo
 \fi
}%
\providecommand \@ifx [1]{%
 \ifx #1\expandafter \@firstoftwo
 \else \expandafter \@secondoftwo
 \fi
}%
\providecommand \natexlab [1]{#1}%
\providecommand \enquote  [1]{``#1''}%
\providecommand \bibnamefont  [1]{#1}%
\providecommand \bibfnamefont [1]{#1}%
\providecommand \citenamefont [1]{#1}%
\providecommand \href@noop [0]{\@secondoftwo}%
\providecommand \href [0]{\begingroup \@sanitize@url \@href}%
\providecommand \@href[1]{\@@startlink{#1}\@@href}%
\providecommand \@@href[1]{\endgroup#1\@@endlink}%
\providecommand \@sanitize@url [0]{\catcode `\\12\catcode `\$12\catcode
  `\&12\catcode `\#12\catcode `\^12\catcode `\_12\catcode `\%12\relax}%
\providecommand \@@startlink[1]{}%
\providecommand \@@endlink[0]{}%
\providecommand \url  [0]{\begingroup\@sanitize@url \@url }%
\providecommand \@url [1]{\endgroup\@href {#1}{\urlprefix }}%
\providecommand \urlprefix  [0]{URL }%
\providecommand \Eprint [0]{\href }%
\providecommand \doibase [0]{http://dx.doi.org/}%
\providecommand \selectlanguage [0]{\@gobble}%
\providecommand \bibinfo  [0]{\@secondoftwo}%
\providecommand \bibfield  [0]{\@secondoftwo}%
\providecommand \translation [1]{[#1]}%
\providecommand \BibitemOpen [0]{}%
\providecommand \bibitemStop [0]{}%
\providecommand \bibitemNoStop [0]{.\EOS\space}%
\providecommand \EOS [0]{\spacefactor3000\relax}%
\providecommand \BibitemShut  [1]{\csname bibitem#1\endcsname}%
\let\auto@bib@innerbib\@empty
\bibitem [{\citenamefont {{\"O}stman}\ \emph
  {et~al.}(2018{\natexlab{a}})\citenamefont {{\"O}stman}, \citenamefont
  {Arnalds}, \citenamefont {Kapaklis}, \citenamefont {Taroni},\ and\
  \citenamefont {Hj{\"o}rvarsson}}]{Ostman_1DIsing_2018}%
  \BibitemOpen
  \bibfield  {author} {\bibinfo {author} {\bibfnamefont {E.}~\bibnamefont
  {{\"O}stman}}, \bibinfo {author} {\bibfnamefont {U.~B.}\ \bibnamefont
  {Arnalds}}, \bibinfo {author} {\bibfnamefont {V.}~\bibnamefont {Kapaklis}},
  \bibinfo {author} {\bibfnamefont {A.}~\bibnamefont {Taroni}}, \ and\ \bibinfo
  {author} {\bibfnamefont {B.}~\bibnamefont {Hj{\"o}rvarsson}},\ }\href
  {\doibase 10.1088/1361-648X/aad0c1} {\bibfield  {journal} {\bibinfo
  {journal} {Journal of Physics: Condensed Matter}\ }\textbf {\bibinfo {volume}
  {30}},\ \bibinfo {pages} {365301} (\bibinfo {year}
  {2018}{\natexlab{a}})}\BibitemShut {NoStop}%
\bibitem [{\citenamefont {Wang}\ \emph {et~al.}(2006)\citenamefont {Wang},
  \citenamefont {Nisoli}, \citenamefont {Freitas}, \citenamefont {Li},
  \citenamefont {McConville}, \citenamefont {Cooley}, \citenamefont {Lund},
  \citenamefont {Samarth}, \citenamefont {Leighton}, \citenamefont {Crespi},\
  and\ \citenamefont {Schiffer}}]{Wang2006}%
  \BibitemOpen
  \bibfield  {author} {\bibinfo {author} {\bibfnamefont {R.~F.}\ \bibnamefont
  {Wang}}, \bibinfo {author} {\bibfnamefont {C.}~\bibnamefont {Nisoli}},
  \bibinfo {author} {\bibfnamefont {R.~S.}\ \bibnamefont {Freitas}}, \bibinfo
  {author} {\bibfnamefont {J.}~\bibnamefont {Li}}, \bibinfo {author}
  {\bibfnamefont {W.}~\bibnamefont {McConville}}, \bibinfo {author}
  {\bibfnamefont {B.~J.}\ \bibnamefont {Cooley}}, \bibinfo {author}
  {\bibfnamefont {M.~S.}\ \bibnamefont {Lund}}, \bibinfo {author}
  {\bibfnamefont {N.}~\bibnamefont {Samarth}}, \bibinfo {author} {\bibfnamefont
  {C.}~\bibnamefont {Leighton}}, \bibinfo {author} {\bibfnamefont {V.~H.}\
  \bibnamefont {Crespi}}, \ and\ \bibinfo {author} {\bibfnamefont
  {P.}~\bibnamefont {Schiffer}},\ }\href {\doibase 10.1038/nature04447}
  {\bibfield  {journal} {\bibinfo  {journal} {Nature (London)}\ }\textbf
  {\bibinfo {volume} {439}},\ \bibinfo {pages} {303} (\bibinfo {year}
  {2006})}\BibitemShut {NoStop}%
\bibitem [{\citenamefont {Nisoli}\ \emph {et~al.}(2013)\citenamefont {Nisoli},
  \citenamefont {Moessner},\ and\ \citenamefont {Schiffer}}]{Nisoli2013}%
  \BibitemOpen
  \bibfield  {author} {\bibinfo {author} {\bibfnamefont {C.}~\bibnamefont
  {Nisoli}}, \bibinfo {author} {\bibfnamefont {R.}~\bibnamefont {Moessner}}, \
  and\ \bibinfo {author} {\bibfnamefont {P.}~\bibnamefont {Schiffer}},\ }\href
  {\doibase 10.1103/RevModPhys.85.1473} {\bibfield  {journal} {\bibinfo
  {journal} {Rev. Mod. Phys.}\ }\textbf {\bibinfo {volume} {85}},\ \bibinfo
  {pages} {1473} (\bibinfo {year} {2013})}\BibitemShut {NoStop}%
\bibitem [{\citenamefont {Heyderman}\ and\ \citenamefont
  {Stamps}(2013)}]{Heyderman:2013gb}%
  \BibitemOpen
  \bibfield  {author} {\bibinfo {author} {\bibfnamefont {L.~J.}\ \bibnamefont
  {Heyderman}}\ and\ \bibinfo {author} {\bibfnamefont {R.~L.}\ \bibnamefont
  {Stamps}},\ }\href {\doibase 10.1088/0953-8984/25/36/363201} {\bibfield
  {journal} {\bibinfo  {journal} {Journal of Physics: Condensed Matter}\
  }\textbf {\bibinfo {volume} {25}},\ \bibinfo {pages} {363201} (\bibinfo
  {year} {2013})}\BibitemShut {NoStop}%
\bibitem [{\citenamefont {Rougemaille}\ and\ \citenamefont
  {Canals}(2019)}]{Rougemaille_Colloquium_2019}%
  \BibitemOpen
  \bibfield  {author} {\bibinfo {author} {\bibfnamefont {N.}~\bibnamefont
  {Rougemaille}}\ and\ \bibinfo {author} {\bibfnamefont {B.}~\bibnamefont
  {Canals}},\ }\href {\doibase 10.1140/ep jb/e2018-90346-7} {\bibfield
  {journal} {\bibinfo  {journal} {Eur. Phys. J. B}\ }\textbf {\bibinfo {volume}
  {92}},\ \bibinfo {pages} {62} (\bibinfo {year} {2019})}\BibitemShut {NoStop}%
\bibitem [{\citenamefont {Kapaklis}\ \emph {et~al.}(2014)\citenamefont
  {Kapaklis}, \citenamefont {Arnalds}, \citenamefont {Farhan}, \citenamefont
  {Chopdekar}, \citenamefont {Balan}, \citenamefont {Scholl}, \citenamefont
  {Heyderman},\ and\ \citenamefont {Hj\"orvarsson}}]{ThermalFluctuationsInASI}%
  \BibitemOpen
  \bibfield  {author} {\bibinfo {author} {\bibfnamefont {V.}~\bibnamefont
  {Kapaklis}}, \bibinfo {author} {\bibfnamefont {U.~B.}\ \bibnamefont
  {Arnalds}}, \bibinfo {author} {\bibfnamefont {A.}~\bibnamefont {Farhan}},
  \bibinfo {author} {\bibfnamefont {R.~V.}\ \bibnamefont {Chopdekar}}, \bibinfo
  {author} {\bibfnamefont {A.}~\bibnamefont {Balan}}, \bibinfo {author}
  {\bibfnamefont {A.}~\bibnamefont {Scholl}}, \bibinfo {author} {\bibfnamefont
  {L.~J.}\ \bibnamefont {Heyderman}}, \ and\ \bibinfo {author} {\bibfnamefont
  {B.}~\bibnamefont {Hj\"orvarsson}},\ }\href {\doibase 10.1038/nnano.2014.104}
  {\bibfield  {journal} {\bibinfo  {journal} {Nat. Nanotechnol.}\ }\textbf
  {\bibinfo {volume} {9}},\ \bibinfo {pages} {514–519} (\bibinfo {year}
  {2014})}\BibitemShut {NoStop}%
\bibitem [{\citenamefont {Perrin}\ \emph {et~al.}(2016)\citenamefont {Perrin},
  \citenamefont {Canals},\ and\ \citenamefont
  {Rougemaille}}]{Perrin_Nature_2016}%
  \BibitemOpen
  \bibfield  {author} {\bibinfo {author} {\bibfnamefont {Y.}~\bibnamefont
  {Perrin}}, \bibinfo {author} {\bibfnamefont {B.}~\bibnamefont {Canals}}, \
  and\ \bibinfo {author} {\bibfnamefont {N.}~\bibnamefont {Rougemaille}},\
  }\href {\doibase 10.1038/nature20155} {\bibfield  {journal} {\bibinfo
  {journal} {Nature}\ }\textbf {\bibinfo {volume} {540}},\ \bibinfo {pages}
  {410} (\bibinfo {year} {2016})}\BibitemShut {NoStop}%
\bibitem [{\citenamefont {{\"O}stman}\ \emph
  {et~al.}(2018{\natexlab{b}})\citenamefont {{\"O}stman}, \citenamefont
  {Stopfel}, \citenamefont {Chioar}, \citenamefont {Arnalds}, \citenamefont
  {Stein}, \citenamefont {Kapaklis},\ and\ \citenamefont
  {Hj{\"o}rvarsson}}]{Ostman_natphys_2018}%
  \BibitemOpen
  \bibfield  {author} {\bibinfo {author} {\bibfnamefont {E.}~\bibnamefont
  {{\"O}stman}}, \bibinfo {author} {\bibfnamefont {H.}~\bibnamefont {Stopfel}},
  \bibinfo {author} {\bibfnamefont {I.-A.}\ \bibnamefont {Chioar}}, \bibinfo
  {author} {\bibfnamefont {U.~B.}\ \bibnamefont {Arnalds}}, \bibinfo {author}
  {\bibfnamefont {A.}~\bibnamefont {Stein}}, \bibinfo {author} {\bibfnamefont
  {V.}~\bibnamefont {Kapaklis}}, \ and\ \bibinfo {author} {\bibfnamefont
  {B.}~\bibnamefont {Hj{\"o}rvarsson}},\ }\href {\doibase
  10.1038/s41567-017-0027-2} {\bibfield  {journal} {\bibinfo  {journal} {Nature
  Physics}\ }\textbf {\bibinfo {volume} {14}},\ \bibinfo {pages} {375}
  (\bibinfo {year} {2018}{\natexlab{b}})}\BibitemShut {NoStop}%
\bibitem [{\citenamefont {Nisoli}\ \emph {et~al.}(2017)\citenamefont {Nisoli},
  \citenamefont {Kapaklis},\ and\ \citenamefont
  {Schiffer}}]{Nisoli_NatPhys_Perspective}%
  \BibitemOpen
  \bibfield  {author} {\bibinfo {author} {\bibfnamefont {C.}~\bibnamefont
  {Nisoli}}, \bibinfo {author} {\bibfnamefont {V.}~\bibnamefont {Kapaklis}}, \
  and\ \bibinfo {author} {\bibfnamefont {P.}~\bibnamefont {Schiffer}},\ }\href
  {\doibase 10.1038/nphys4059} {\bibfield  {journal} {\bibinfo  {journal}
  {Nature Physics}\ }\textbf {\bibinfo {volume} {13}},\ \bibinfo {pages} {200}
  (\bibinfo {year} {2017})}\BibitemShut {NoStop}%
\bibitem [{\citenamefont {Pohlit}\ \emph {et~al.}(2015)\citenamefont {Pohlit},
  \citenamefont {Porrati}, \citenamefont {Huth}, \citenamefont {Ohno},
  \citenamefont {Ohno},\ and\ \citenamefont {M{\"u}ller}}]{Pohlit_JAP_2015}%
  \BibitemOpen
  \bibfield  {author} {\bibinfo {author} {\bibfnamefont {M.}~\bibnamefont
  {Pohlit}}, \bibinfo {author} {\bibfnamefont {F.}~\bibnamefont {Porrati}},
  \bibinfo {author} {\bibfnamefont {M.}~\bibnamefont {Huth}}, \bibinfo {author}
  {\bibfnamefont {Y.}~\bibnamefont {Ohno}}, \bibinfo {author} {\bibfnamefont
  {H.}~\bibnamefont {Ohno}}, \ and\ \bibinfo {author} {\bibfnamefont
  {J.}~\bibnamefont {M{\"u}ller}},\ }\href {\doibase 10.1063/1.4917497}
  {\bibfield  {journal} {\bibinfo  {journal} {Journal of Applied Physics}\
  }\textbf {\bibinfo {volume} {117}},\ \bibinfo {pages} {17C746} (\bibinfo
  {year} {2015})}\BibitemShut {NoStop}%
\bibitem [{\citenamefont {Pohlit}\ \emph {et~al.}(2016)\citenamefont {Pohlit},
  \citenamefont {Stockem}, \citenamefont {Porrati}, \citenamefont {Huth},
  \citenamefont {Schr{\"o}der},\ and\ \citenamefont
  {M{\"u}ller}}]{Pohlit_JAP_2016}%
  \BibitemOpen
  \bibfield  {author} {\bibinfo {author} {\bibfnamefont {M.}~\bibnamefont
  {Pohlit}}, \bibinfo {author} {\bibfnamefont {I.}~\bibnamefont {Stockem}},
  \bibinfo {author} {\bibfnamefont {F.}~\bibnamefont {Porrati}}, \bibinfo
  {author} {\bibfnamefont {M.}~\bibnamefont {Huth}}, \bibinfo {author}
  {\bibfnamefont {C.}~\bibnamefont {Schr{\"o}der}}, \ and\ \bibinfo {author}
  {\bibfnamefont {J.}~\bibnamefont {M{\"u}ller}},\ }\href {\doibase
  10.1063/1.4961705} {\bibfield  {journal} {\bibinfo  {journal} {Journal of
  Applied Physics}\ }\textbf {\bibinfo {volume} {120}},\ \bibinfo {pages}
  {142103} (\bibinfo {year} {2016})}\BibitemShut {NoStop}%
\bibitem [{\citenamefont {Kapaklis}\ \emph {et~al.}(2012)\citenamefont
  {Kapaklis}, \citenamefont {Arnalds}, \citenamefont {Harman-Clarke},
  \citenamefont {Papaioannou}, \citenamefont {Karimipour}, \citenamefont
  {Korelis}, \citenamefont {Taroni}, \citenamefont {Holdsworth}, \citenamefont
  {Bramwell},\ and\ \citenamefont
  {Hj{\"o}rvarsson}}]{melting_artificial_spin_ice}%
  \BibitemOpen
  \bibfield  {author} {\bibinfo {author} {\bibfnamefont {V.}~\bibnamefont
  {Kapaklis}}, \bibinfo {author} {\bibfnamefont {U.~B.}\ \bibnamefont
  {Arnalds}}, \bibinfo {author} {\bibfnamefont {A.}~\bibnamefont
  {Harman-Clarke}}, \bibinfo {author} {\bibfnamefont {E.~T.}\ \bibnamefont
  {Papaioannou}}, \bibinfo {author} {\bibfnamefont {M.}~\bibnamefont
  {Karimipour}}, \bibinfo {author} {\bibfnamefont {P.}~\bibnamefont {Korelis}},
  \bibinfo {author} {\bibfnamefont {A.}~\bibnamefont {Taroni}}, \bibinfo
  {author} {\bibfnamefont {P.~C.~W.}\ \bibnamefont {Holdsworth}}, \bibinfo
  {author} {\bibfnamefont {S.~T.}\ \bibnamefont {Bramwell}}, \ and\ \bibinfo
  {author} {\bibfnamefont {B.}~\bibnamefont {Hj{\"o}rvarsson}},\ }\href
  {http://stacks.iop.org/1367-2630/14/i=3/a=035009} {\bibfield  {journal}
  {\bibinfo  {journal} {New Journal of Physics}\ }\textbf {\bibinfo {volume}
  {14}},\ \bibinfo {pages} {035009} (\bibinfo {year} {2012})}\BibitemShut
  {NoStop}%
\bibitem [{\citenamefont {Morgan}\ \emph {et~al.}(2010)\citenamefont {Morgan},
  \citenamefont {Stein}, \citenamefont {Langridge},\ and\ \citenamefont
  {Marrows}}]{Morgan_natphys_2010}%
  \BibitemOpen
  \bibfield  {author} {\bibinfo {author} {\bibfnamefont {J.~P.}\ \bibnamefont
  {Morgan}}, \bibinfo {author} {\bibfnamefont {A.}~\bibnamefont {Stein}},
  \bibinfo {author} {\bibfnamefont {S.}~\bibnamefont {Langridge}}, \ and\
  \bibinfo {author} {\bibfnamefont {C.~H.}\ \bibnamefont {Marrows}},\ }\href
  {\doibase 10.1038/nphys1853} {\bibfield  {journal} {\bibinfo  {journal}
  {Nature Physics}\ }\textbf {\bibinfo {volume} {7}},\ \bibinfo {pages} {75}
  (\bibinfo {year} {2010})}\BibitemShut {NoStop}%
\bibitem [{\citenamefont {Farhan}\ \emph {et~al.}(2013)\citenamefont {Farhan},
  \citenamefont {Derlet}, \citenamefont {Kleibert}, \citenamefont {Balan},
  \citenamefont {Chopdekar}, \citenamefont {Wyss}, \citenamefont {Anghinolfi},
  \citenamefont {Nolting},\ and\ \citenamefont {Heyderman}}]{Farhan:2013ki}%
  \BibitemOpen
  \bibfield  {author} {\bibinfo {author} {\bibfnamefont {A.}~\bibnamefont
  {Farhan}}, \bibinfo {author} {\bibfnamefont {P.~M.}\ \bibnamefont {Derlet}},
  \bibinfo {author} {\bibfnamefont {A.}~\bibnamefont {Kleibert}}, \bibinfo
  {author} {\bibfnamefont {A.}~\bibnamefont {Balan}}, \bibinfo {author}
  {\bibfnamefont {R.~V.}\ \bibnamefont {Chopdekar}}, \bibinfo {author}
  {\bibfnamefont {M.}~\bibnamefont {Wyss}}, \bibinfo {author} {\bibfnamefont
  {L.}~\bibnamefont {Anghinolfi}}, \bibinfo {author} {\bibfnamefont
  {F.}~\bibnamefont {Nolting}}, \ and\ \bibinfo {author} {\bibfnamefont
  {L.~J.}\ \bibnamefont {Heyderman}},\ }\href {\doibase 10.1038/nphys2613}
  {\bibfield  {journal} {\bibinfo  {journal} {Nature Physics}\ }\textbf
  {\bibinfo {volume} {9}},\ \bibinfo {pages} {375} (\bibinfo {year}
  {2013})}\BibitemShut {NoStop}%
\bibitem [{\citenamefont {Porro}\ \emph {et~al.}(2013)\citenamefont {Porro},
  \citenamefont {Bedoya-Pinto}, \citenamefont {Berger},\ and\ \citenamefont
  {Vavassori}}]{Porro:2013cm}%
  \BibitemOpen
  \bibfield  {author} {\bibinfo {author} {\bibfnamefont {J.~M.}\ \bibnamefont
  {Porro}}, \bibinfo {author} {\bibfnamefont {A.}~\bibnamefont {Bedoya-Pinto}},
  \bibinfo {author} {\bibfnamefont {A.}~\bibnamefont {Berger}}, \ and\ \bibinfo
  {author} {\bibfnamefont {P.}~\bibnamefont {Vavassori}},\ }\href {\doibase
  10.1088/1367-2630/15/5/055012} {\bibfield  {journal} {\bibinfo  {journal}
  {New Journal of Physics}\ }\textbf {\bibinfo {volume} {15}},\ \bibinfo
  {pages} {055012} (\bibinfo {year} {2013})}\BibitemShut {NoStop}%
\bibitem [{\citenamefont {Andersson}\ \emph {et~al.}(2016)\citenamefont
  {Andersson}, \citenamefont {Pappas}, \citenamefont {Stopfel}, \citenamefont
  {\"Ostman}, \citenamefont {Stein}, \citenamefont {Nordblad}, \citenamefont
  {Mathieu}, \citenamefont {Hj\"orvarsson},\ and\ \citenamefont
  {Kapaklis}}]{Andersson2016}%
  \BibitemOpen
  \bibfield  {author} {\bibinfo {author} {\bibfnamefont {M.~S.}\ \bibnamefont
  {Andersson}}, \bibinfo {author} {\bibfnamefont {S.~D.}\ \bibnamefont
  {Pappas}}, \bibinfo {author} {\bibfnamefont {H.}~\bibnamefont {Stopfel}},
  \bibinfo {author} {\bibfnamefont {E.}~\bibnamefont {\"Ostman}}, \bibinfo
  {author} {\bibfnamefont {A.}~\bibnamefont {Stein}}, \bibinfo {author}
  {\bibfnamefont {P.}~\bibnamefont {Nordblad}}, \bibinfo {author}
  {\bibfnamefont {R.}~\bibnamefont {Mathieu}}, \bibinfo {author} {\bibfnamefont
  {B.}~\bibnamefont {Hj\"orvarsson}}, \ and\ \bibinfo {author} {\bibfnamefont
  {V.}~\bibnamefont {Kapaklis}},\ }\href {\doibase 10.1038/srep37097}
  {\bibfield  {journal} {\bibinfo  {journal} {Scientific Reports}\ }\textbf
  {\bibinfo {volume} {6}},\ \bibinfo {pages} {37097} (\bibinfo {year}
  {2016})}\BibitemShut {NoStop}%
\bibitem [{\citenamefont {Morley}\ \emph {et~al.}(2017)\citenamefont {Morley},
  \citenamefont {Venero}, \citenamefont {Porro}, \citenamefont {Riley},
  \citenamefont {Stein}, \citenamefont {Steadman}, \citenamefont {Stamps},
  \citenamefont {Langridge},\ and\ \citenamefont {Marrows}}]{Morley2017}%
  \BibitemOpen
  \bibfield  {author} {\bibinfo {author} {\bibfnamefont {S.~A.}\ \bibnamefont
  {Morley}}, \bibinfo {author} {\bibfnamefont {D.~A.}\ \bibnamefont {Venero}},
  \bibinfo {author} {\bibfnamefont {J.~M.}\ \bibnamefont {Porro}}, \bibinfo
  {author} {\bibfnamefont {S.~T.}\ \bibnamefont {Riley}}, \bibinfo {author}
  {\bibfnamefont {A.}~\bibnamefont {Stein}}, \bibinfo {author} {\bibfnamefont
  {P.}~\bibnamefont {Steadman}}, \bibinfo {author} {\bibfnamefont {R.~L.}\
  \bibnamefont {Stamps}}, \bibinfo {author} {\bibfnamefont {S.}~\bibnamefont
  {Langridge}}, \ and\ \bibinfo {author} {\bibfnamefont {C.~H.}\ \bibnamefont
  {Marrows}},\ }\href {\doibase 10.1103/physrevb.95.104422} {\bibfield
  {journal} {\bibinfo  {journal} {Physical Review B}\ }\textbf {\bibinfo
  {volume} {95}},\ \bibinfo {pages} {104422} (\bibinfo {year}
  {2017})}\BibitemShut {NoStop}%
\bibitem [{\citenamefont {Sendetskyi}\ \emph {et~al.}(2019)\citenamefont
  {Sendetskyi}, \citenamefont {Scagnoli}, \citenamefont {Leo}, \citenamefont
  {Anghinolfi}, \citenamefont {Alberca}, \citenamefont {Lüning}, \citenamefont
  {Staub}, \citenamefont {Derlet},\ and\ \citenamefont
  {Heyderman}}]{Sendetskyi2019}%
  \BibitemOpen
  \bibfield  {author} {\bibinfo {author} {\bibfnamefont {O.}~\bibnamefont
  {Sendetskyi}}, \bibinfo {author} {\bibfnamefont {V.}~\bibnamefont
  {Scagnoli}}, \bibinfo {author} {\bibfnamefont {N.}~\bibnamefont {Leo}},
  \bibinfo {author} {\bibfnamefont {L.}~\bibnamefont {Anghinolfi}}, \bibinfo
  {author} {\bibfnamefont {A.}~\bibnamefont {Alberca}}, \bibinfo {author}
  {\bibfnamefont {J.}~\bibnamefont {Lüning}}, \bibinfo {author} {\bibfnamefont
  {U.}~\bibnamefont {Staub}}, \bibinfo {author} {\bibfnamefont {P.~M.}\
  \bibnamefont {Derlet}}, \ and\ \bibinfo {author} {\bibfnamefont {L.~J.}\
  \bibnamefont {Heyderman}},\ }\href {\doibase 10.1103/physrevb.99.214430}
  {\bibfield  {journal} {\bibinfo  {journal} {Physical Review B}\ }\textbf
  {\bibinfo {volume} {99}},\ \bibinfo {pages} {214430} (\bibinfo {year}
  {2019})}\BibitemShut {NoStop}%
\bibitem [{\citenamefont {Saccone}\ \emph {et~al.}(2019)\citenamefont
  {Saccone}, \citenamefont {Scholl}, \citenamefont {Velten}, \citenamefont
  {Dhuey}, \citenamefont {Hofhuis}, \citenamefont {Wuth}, \citenamefont
  {Huang}, \citenamefont {Chen}, \citenamefont {Chopdekar},\ and\ \citenamefont
  {Farhan}}]{Saccone2019}%
  \BibitemOpen
  \bibfield  {author} {\bibinfo {author} {\bibfnamefont {M.}~\bibnamefont
  {Saccone}}, \bibinfo {author} {\bibfnamefont {A.}~\bibnamefont {Scholl}},
  \bibinfo {author} {\bibfnamefont {S.}~\bibnamefont {Velten}}, \bibinfo
  {author} {\bibfnamefont {S.}~\bibnamefont {Dhuey}}, \bibinfo {author}
  {\bibfnamefont {K.}~\bibnamefont {Hofhuis}}, \bibinfo {author} {\bibfnamefont
  {C.}~\bibnamefont {Wuth}}, \bibinfo {author} {\bibfnamefont {Y.-L.}\
  \bibnamefont {Huang}}, \bibinfo {author} {\bibfnamefont {Z.}~\bibnamefont
  {Chen}}, \bibinfo {author} {\bibfnamefont {R.~V.}\ \bibnamefont {Chopdekar}},
  \ and\ \bibinfo {author} {\bibfnamefont {A.}~\bibnamefont {Farhan}},\ }\href
  {\doibase 10.1103/physrevb.99.224403} {\bibfield  {journal} {\bibinfo
  {journal} {Physical Review B}\ }\textbf {\bibinfo {volume} {99}},\ \bibinfo
  {pages} {224403} (\bibinfo {year} {2019})}\BibitemShut {NoStop}%
\bibitem [{\citenamefont {Anghinolfi}\ \emph {et~al.}(2015)\citenamefont
  {Anghinolfi}, \citenamefont {Luetkens}, \citenamefont {Perron}, \citenamefont
  {Flokstra}, \citenamefont {Sendetskyi}, \citenamefont {Suter}, \citenamefont
  {Prokscha}, \citenamefont {Derlet}, \citenamefont {Lee},\ and\ \citenamefont
  {Heyderman}}]{LJHeyderman}%
  \BibitemOpen
  \bibfield  {author} {\bibinfo {author} {\bibfnamefont {L.}~\bibnamefont
  {Anghinolfi}}, \bibinfo {author} {\bibfnamefont {H.}~\bibnamefont
  {Luetkens}}, \bibinfo {author} {\bibfnamefont {J.}~\bibnamefont {Perron}},
  \bibinfo {author} {\bibfnamefont {M.~G.}\ \bibnamefont {Flokstra}}, \bibinfo
  {author} {\bibfnamefont {O.}~\bibnamefont {Sendetskyi}}, \bibinfo {author}
  {\bibfnamefont {A.}~\bibnamefont {Suter}}, \bibinfo {author} {\bibfnamefont
  {T.}~\bibnamefont {Prokscha}}, \bibinfo {author} {\bibfnamefont {P.~M.}\
  \bibnamefont {Derlet}}, \bibinfo {author} {\bibfnamefont {S.~L.}\
  \bibnamefont {Lee}}, \ and\ \bibinfo {author} {\bibfnamefont {L.~J.}\
  \bibnamefont {Heyderman}},\ }\href {\doibase 10.1038/ncomms9278} {\bibfield
  {journal} {\bibinfo  {journal} {Nat. Commun.}\ }\textbf {\bibinfo {volume}
  {6}},\ \bibinfo {pages} {8278} (\bibinfo {year} {2015})}\BibitemShut
  {NoStop}%
\bibitem [{\citenamefont {Leo}\ \emph {et~al.}(2018)\citenamefont {Leo},
  \citenamefont {Holenstein}, \citenamefont {Schildknecht}, \citenamefont
  {Sendetskyi}, \citenamefont {Luetkens}, \citenamefont {Derlet}, \citenamefont
  {Scagnoli}, \citenamefont {Lan{\c c}on}, \citenamefont {Mardegan},
  \citenamefont {Prokscha}, \citenamefont {Suter}, \citenamefont {Salman},
  \citenamefont {Lee},\ and\ \citenamefont {Heyderman}}]{Leo:2018di}%
  \BibitemOpen
  \bibfield  {author} {\bibinfo {author} {\bibfnamefont {N.}~\bibnamefont
  {Leo}}, \bibinfo {author} {\bibfnamefont {S.}~\bibnamefont {Holenstein}},
  \bibinfo {author} {\bibfnamefont {D.}~\bibnamefont {Schildknecht}}, \bibinfo
  {author} {\bibfnamefont {O.}~\bibnamefont {Sendetskyi}}, \bibinfo {author}
  {\bibfnamefont {H.}~\bibnamefont {Luetkens}}, \bibinfo {author}
  {\bibfnamefont {P.~M.}\ \bibnamefont {Derlet}}, \bibinfo {author}
  {\bibfnamefont {V.}~\bibnamefont {Scagnoli}}, \bibinfo {author}
  {\bibfnamefont {D.}~\bibnamefont {Lan{\c c}on}}, \bibinfo {author}
  {\bibfnamefont {J.~R.~L.}\ \bibnamefont {Mardegan}}, \bibinfo {author}
  {\bibfnamefont {T.}~\bibnamefont {Prokscha}}, \bibinfo {author}
  {\bibfnamefont {A.}~\bibnamefont {Suter}}, \bibinfo {author} {\bibfnamefont
  {Z.}~\bibnamefont {Salman}}, \bibinfo {author} {\bibfnamefont
  {S.}~\bibnamefont {Lee}}, \ and\ \bibinfo {author} {\bibfnamefont {L.~J.}\
  \bibnamefont {Heyderman}},\ }\href {\doibase 10.1038/s41467-018-05216-2}
  {\bibfield  {journal} {\bibinfo  {journal} {Nature Communications}\ }\textbf
  {\bibinfo {volume} {9}},\ \bibinfo {pages} {2850} (\bibinfo {year}
  {2018})}\BibitemShut {NoStop}%
\bibitem [{\citenamefont {Bedanta}\ and\ \citenamefont
  {Kleemann}(2008)}]{Bedanta2008}%
  \BibitemOpen
  \bibfield  {author} {\bibinfo {author} {\bibfnamefont {S.}~\bibnamefont
  {Bedanta}}\ and\ \bibinfo {author} {\bibfnamefont {W.}~\bibnamefont
  {Kleemann}},\ }\href {\doibase 10.1088/0022-3727/42/1/013001} {\bibfield
  {journal} {\bibinfo  {journal} {Journal of Physics D: Applied Physics}\
  }\textbf {\bibinfo {volume} {42}},\ \bibinfo {pages} {013001} (\bibinfo
  {year} {2008})}\BibitemShut {NoStop}%
\bibitem [{\citenamefont {Topping}\ and\ \citenamefont
  {Blundell}(2018)}]{Topping_2019_review}%
  \BibitemOpen
  \bibfield  {author} {\bibinfo {author} {\bibfnamefont {C.~V.}\ \bibnamefont
  {Topping}}\ and\ \bibinfo {author} {\bibfnamefont {S.~J.}\ \bibnamefont
  {Blundell}},\ }\href {\doibase 10.1088/1361-648X/aaed96} {\bibfield
  {journal} {\bibinfo  {journal} {Journal of Physics: Condensed Matter}\
  }\textbf {\bibinfo {volume} {31}},\ \bibinfo {pages} {013001} (\bibinfo
  {year} {2018})}\BibitemShut {NoStop}%
\bibitem [{\citenamefont {Landi}(2013)}]{LandiJAP2013}%
  \BibitemOpen
  \bibfield  {author} {\bibinfo {author} {\bibfnamefont {G.~T.}\ \bibnamefont
  {Landi}},\ }\href {\doibase 10.1063/1.4802583} {\bibfield  {journal}
  {\bibinfo  {journal} {Journal of Applied Physics}\ }\textbf {\bibinfo
  {volume} {113}},\ \bibinfo {pages} {163908} (\bibinfo {year}
  {2013})}\BibitemShut {NoStop}%
\bibitem [{\citenamefont {Vernay}\ \emph {et~al.}(2014)\citenamefont {Vernay},
  \citenamefont {Sabsabi},\ and\ \citenamefont {Kachkachi}}]{Vernay_PRB_2014}%
  \BibitemOpen
  \bibfield  {author} {\bibinfo {author} {\bibfnamefont {F.}~\bibnamefont
  {Vernay}}, \bibinfo {author} {\bibfnamefont {Z.}~\bibnamefont {Sabsabi}}, \
  and\ \bibinfo {author} {\bibfnamefont {H.}~\bibnamefont {Kachkachi}},\ }\href
  {\doibase 10.1103/PhysRevB.90.094416} {\bibfield  {journal} {\bibinfo
  {journal} {Phys. Rev. B}\ }\textbf {\bibinfo {volume} {90}},\ \bibinfo
  {pages} {094416} (\bibinfo {year} {2014})}\BibitemShut {NoStop}%
\bibitem [{\citenamefont {P{\"a}rnaste}\ \emph {et~al.}(2007)\citenamefont
  {P{\"a}rnaste}, \citenamefont {Marcellini}, \citenamefont {Holmstr{\"o}m},
  \citenamefont {Bock}, \citenamefont {Fransson}, \citenamefont {Eriksson},\
  and\ \citenamefont {Hj{\"o}rvarsson}}]{parnaste2007dimensionality}%
  \BibitemOpen
  \bibfield  {author} {\bibinfo {author} {\bibfnamefont {M.}~\bibnamefont
  {P{\"a}rnaste}}, \bibinfo {author} {\bibfnamefont {M.}~\bibnamefont
  {Marcellini}}, \bibinfo {author} {\bibfnamefont {E.}~\bibnamefont
  {Holmstr{\"o}m}}, \bibinfo {author} {\bibfnamefont {N.}~\bibnamefont {Bock}},
  \bibinfo {author} {\bibfnamefont {J.}~\bibnamefont {Fransson}}, \bibinfo
  {author} {\bibfnamefont {O.}~\bibnamefont {Eriksson}}, \ and\ \bibinfo
  {author} {\bibfnamefont {B.}~\bibnamefont {Hj{\"o}rvarsson}},\ }\href
  {\doibase 10.1088/0953-8984/19/24/246213} {\bibfield  {journal} {\bibinfo
  {journal} {Journal of Physics: Condensed Matter}\ }\textbf {\bibinfo {volume}
  {19}},\ \bibinfo {pages} {246213} (\bibinfo {year} {2007})}\BibitemShut
  {NoStop}%
\bibitem [{\citenamefont {Hase}\ \emph {et~al.}(2014)\citenamefont {Hase},
  \citenamefont {Brewer}, \citenamefont {Arnalds}, \citenamefont {Ahlberg},
  \citenamefont {Kapaklis}, \citenamefont {Bj{\"o}rck}, \citenamefont
  {Bouchenoire}, \citenamefont {Thompson}, \citenamefont {Haskel},
  \citenamefont {Choi}, \citenamefont {Lang}, \citenamefont
  {S{\'a}nchez-Hanke},\ and\ \citenamefont
  {Hj{\"o}rvarsson}}]{Hase_PRB_Delta_polarizability_2014}%
  \BibitemOpen
  \bibfield  {author} {\bibinfo {author} {\bibfnamefont {T.~P.~A.}\
  \bibnamefont {Hase}}, \bibinfo {author} {\bibfnamefont {M.~S.}\ \bibnamefont
  {Brewer}}, \bibinfo {author} {\bibfnamefont {U.~B.}\ \bibnamefont {Arnalds}},
  \bibinfo {author} {\bibfnamefont {M.}~\bibnamefont {Ahlberg}}, \bibinfo
  {author} {\bibfnamefont {V.}~\bibnamefont {Kapaklis}}, \bibinfo {author}
  {\bibfnamefont {M.}~\bibnamefont {Bj{\"o}rck}}, \bibinfo {author}
  {\bibfnamefont {L.}~\bibnamefont {Bouchenoire}}, \bibinfo {author}
  {\bibfnamefont {P.}~\bibnamefont {Thompson}}, \bibinfo {author}
  {\bibfnamefont {D.}~\bibnamefont {Haskel}}, \bibinfo {author} {\bibfnamefont
  {Y.}~\bibnamefont {Choi}}, \bibinfo {author} {\bibfnamefont {J.}~\bibnamefont
  {Lang}}, \bibinfo {author} {\bibfnamefont {C.}~\bibnamefont
  {S{\'a}nchez-Hanke}}, \ and\ \bibinfo {author} {\bibfnamefont
  {B.}~\bibnamefont {Hj{\"o}rvarsson}},\ }\href {\doibase
  10.1103/PhysRevB.90.104403} {\bibfield  {journal} {\bibinfo  {journal}
  {Physical Review B}\ }\textbf {\bibinfo {volume} {90}},\ \bibinfo {pages}
  {104403} (\bibinfo {year} {2014})}\BibitemShut {NoStop}%
\bibitem [{Note1()}]{Note1}%
  \BibitemOpen
  \bibinfo {note} {The small difference between the moment stated here compared
  to \protect \citet {Andersson2016} stems from a more detailed analyses of the
  island dimensions from SEM imaging.}\BibitemShut {Stop}%
\bibitem [{\citenamefont {Vansteenkiste}\ \emph {et~al.}(2014)\citenamefont
  {Vansteenkiste}, \citenamefont {Leliaert}, \citenamefont {Dvornik},
  \citenamefont {Helsen}, \citenamefont {Garcia-Sanchez},\ and\ \citenamefont
  {Waeyenberge}}]{Vansteenkiste2014}%
  \BibitemOpen
  \bibfield  {author} {\bibinfo {author} {\bibfnamefont {A.}~\bibnamefont
  {Vansteenkiste}}, \bibinfo {author} {\bibfnamefont {J.}~\bibnamefont
  {Leliaert}}, \bibinfo {author} {\bibfnamefont {M.}~\bibnamefont {Dvornik}},
  \bibinfo {author} {\bibfnamefont {M.}~\bibnamefont {Helsen}}, \bibinfo
  {author} {\bibfnamefont {F.}~\bibnamefont {Garcia-Sanchez}}, \ and\ \bibinfo
  {author} {\bibfnamefont {B.~V.}\ \bibnamefont {Waeyenberge}},\ }\href
  {\doibase 10.1063/1.4899186} {\bibfield  {journal} {\bibinfo  {journal}
  {{AIP} Advances}\ }\textbf {\bibinfo {volume} {4}},\ \bibinfo {pages}
  {107133} (\bibinfo {year} {2014})}\BibitemShut {NoStop}%
\bibitem [{\citenamefont {Bessarab}\ \emph {et~al.}(2012)\citenamefont
  {Bessarab}, \citenamefont {Uzdin},\ and\ \citenamefont
  {J{\'{o}}nsson}}]{Bessarab2012}%
  \BibitemOpen
  \bibfield  {author} {\bibinfo {author} {\bibfnamefont {P.~F.}\ \bibnamefont
  {Bessarab}}, \bibinfo {author} {\bibfnamefont {V.~M.}\ \bibnamefont {Uzdin}},
  \ and\ \bibinfo {author} {\bibfnamefont {H.}~\bibnamefont {J{\'{o}}nsson}},\
  }\href {\doibase 10.1103/physrevb.85.184409} {\bibfield  {journal} {\bibinfo
  {journal} {Physical Review B}\ }\textbf {\bibinfo {volume} {85}},\ \bibinfo
  {pages} {184409} (\bibinfo {year} {2012})}\BibitemShut {NoStop}%
\bibitem [{\citenamefont {Bessarab}\ \emph {et~al.}(2013)\citenamefont
  {Bessarab}, \citenamefont {Uzdin},\ and\ \citenamefont
  {J{\'{o}}nsson}}]{Bessarab2013}%
  \BibitemOpen
  \bibfield  {author} {\bibinfo {author} {\bibfnamefont {P.~F.}\ \bibnamefont
  {Bessarab}}, \bibinfo {author} {\bibfnamefont {V.~M.}\ \bibnamefont {Uzdin}},
  \ and\ \bibinfo {author} {\bibfnamefont {H.}~\bibnamefont {J{\'{o}}nsson}},\
  }\href {\doibase 10.1103/physrevlett.110.020604} {\bibfield  {journal}
  {\bibinfo  {journal} {Physical Review Letters}\ }\textbf {\bibinfo {volume}
  {110}},\ \bibinfo {pages} {020604} (\bibinfo {year} {2013})}\BibitemShut
  {NoStop}%
\bibitem [{\citenamefont {Gliga}\ \emph {et~al.}(2015)\citenamefont {Gliga},
  \citenamefont {K{\'a}kay}, \citenamefont {Heyderman}, \citenamefont
  {Hertel},\ and\ \citenamefont {Heinonen}}]{Gliga_PRB_2015}%
  \BibitemOpen
  \bibfield  {author} {\bibinfo {author} {\bibfnamefont {S.}~\bibnamefont
  {Gliga}}, \bibinfo {author} {\bibfnamefont {A.}~\bibnamefont {K{\'a}kay}},
  \bibinfo {author} {\bibfnamefont {L.~J.}\ \bibnamefont {Heyderman}}, \bibinfo
  {author} {\bibfnamefont {R.}~\bibnamefont {Hertel}}, \ and\ \bibinfo {author}
  {\bibfnamefont {O.~G.}\ \bibnamefont {Heinonen}},\ }\href {\doibase
  10.1103/PhysRevB.92.060413} {\bibfield  {journal} {\bibinfo  {journal}
  {Physical Review B}\ }\textbf {\bibinfo {volume} {92}},\ \bibinfo {pages}
  {060413} (\bibinfo {year} {2015})}\BibitemShut {NoStop}%
\bibitem [{\citenamefont {Aspelmeier}\ \emph {et~al.}(1995)\citenamefont
  {Aspelmeier}, \citenamefont {Tischer}, \citenamefont {Farle}, \citenamefont
  {Russo}, \citenamefont {Baberschke},\ and\ \citenamefont
  {Arvanitis}}]{ASPELMEIER_AC_Susceptibility_1995}%
  \BibitemOpen
  \bibfield  {author} {\bibinfo {author} {\bibfnamefont {A.}~\bibnamefont
  {Aspelmeier}}, \bibinfo {author} {\bibfnamefont {M.}~\bibnamefont {Tischer}},
  \bibinfo {author} {\bibfnamefont {M.}~\bibnamefont {Farle}}, \bibinfo
  {author} {\bibfnamefont {M.}~\bibnamefont {Russo}}, \bibinfo {author}
  {\bibfnamefont {K.}~\bibnamefont {Baberschke}}, \ and\ \bibinfo {author}
  {\bibfnamefont {D.}~\bibnamefont {Arvanitis}},\ }\href {\doibase
  https://doi.org/10.1016/0304-8853(95)00025-9} {\bibfield  {journal} {\bibinfo
   {journal} {Journal of Magnetism and Magnetic Materials}\ }\textbf {\bibinfo
  {volume} {146}},\ \bibinfo {pages} {256 } (\bibinfo {year}
  {1995})}\BibitemShut {NoStop}%
\bibitem [{Note2()}]{Note2}%
  \BibitemOpen
  \bibinfo {note} {As temperature gradients between sample surface and
  temperature sensor of the set-up cannot be fully excluded it is plausible to
  assume the presence of temperature drifts of the sample of up to $1\protect
  \tmspace +\thinmuskip {.1667em}\protect \rm {K}$.}\BibitemShut {Stop}%
\bibitem [{\citenamefont {Souletie}\ and\ \citenamefont
  {Tholence}(1985)}]{souletie_critical_1985}%
  \BibitemOpen
  \bibfield  {author} {\bibinfo {author} {\bibfnamefont {J.}~\bibnamefont
  {Souletie}}\ and\ \bibinfo {author} {\bibfnamefont {J.~L.}\ \bibnamefont
  {Tholence}},\ }\href {\doibase 10.1103/PhysRevB.32.516} {\bibfield  {journal}
  {\bibinfo  {journal} {Physical Review B}\ }\textbf {\bibinfo {volume} {32}},\
  \bibinfo {pages} {516} (\bibinfo {year} {1985})}\BibitemShut {NoStop}%
\bibitem [{\citenamefont {Shtrikman}\ and\ \citenamefont
  {Wohlfarth}(1981)}]{Shtrikman1981}%
  \BibitemOpen
  \bibfield  {author} {\bibinfo {author} {\bibfnamefont {S.}~\bibnamefont
  {Shtrikman}}\ and\ \bibinfo {author} {\bibfnamefont {E.}~\bibnamefont
  {Wohlfarth}},\ }\href {\doibase 10.1016/0375-9601(81)90441-2} {\bibfield
  {journal} {\bibinfo  {journal} {Physics Letters A}\ }\textbf {\bibinfo
  {volume} {85}},\ \bibinfo {pages} {467} (\bibinfo {year} {1981})}\BibitemShut
  {NoStop}%
\bibitem [{\citenamefont {Osborn}(1945)}]{PhysRev.67.351}%
  \BibitemOpen
  \bibfield  {author} {\bibinfo {author} {\bibfnamefont {J.~A.}\ \bibnamefont
  {Osborn}},\ }\href {\doibase 10.1103/PhysRev.67.351} {\bibfield  {journal}
  {\bibinfo  {journal} {Phys. Rev.}\ }\textbf {\bibinfo {volume} {67}},\
  \bibinfo {pages} {351} (\bibinfo {year} {1945})}\BibitemShut {NoStop}%
\bibitem [{Note3()}]{Note3}%
  \BibitemOpen
  \bibinfo {note} {Note that, both for the parabola as well as the nonlinear
  VFT-fits, the weighting is defined by the inverse quadratic form of the
  individual errors for each data point, i.e. $w_{\protect \rm {i}}=1/\sigma
  _{\protect \rm {i}}^{2}$. Furthermore, the standard errors of the fitting
  parameters are scaled with the square root of the reduced chi
  squared.}\BibitemShut {Stop}%
\bibitem [{Note4()}]{Note4}%
  \BibitemOpen
  \bibinfo {note} {The micromagnetic simulations have been performed using the
  $\protect \rm {MuMax}^3$ GPU-accelerated micromagnetic simulation program
  \cite {Vansteenkiste2014}. The simulations were performed using the following
  parameters: $M=499289\protect \tmspace +\thinmuskip {.1667em}\protect \rm
  {A/m}$, $A_{\protect \rm {ex}}=3.36\cdot 10^{-12} \protect \rm {J/m}$, length
  $=310\protect \tmspace +\thinmuskip {.1667em}\protect \rm {nm}$, width
  $=140\protect \tmspace +\thinmuskip {.1667em}\protect \rm {nm}$, and a
  thickness of $2\protect \tmspace +\thinmuskip {.1667em}\protect \rm
  {nm}$.}\BibitemShut {Stop}%
\bibitem [{\citenamefont {Gliga}\ \emph {et~al.}(2013)\citenamefont {Gliga},
  \citenamefont {K{\'a}kay}, \citenamefont {Hertel},\ and\ \citenamefont
  {Heinonen}}]{Gliga_PRL_2013}%
  \BibitemOpen
  \bibfield  {author} {\bibinfo {author} {\bibfnamefont {S.}~\bibnamefont
  {Gliga}}, \bibinfo {author} {\bibfnamefont {A.}~\bibnamefont {K{\'a}kay}},
  \bibinfo {author} {\bibfnamefont {R.}~\bibnamefont {Hertel}}, \ and\ \bibinfo
  {author} {\bibfnamefont {O.~G.}\ \bibnamefont {Heinonen}},\ }\href {\doibase
  10.1103/PhysRevLett.110.117205} {\bibfield  {journal} {\bibinfo  {journal}
  {Physical Review Letters}\ }\textbf {\bibinfo {volume} {110}},\ \bibinfo
  {pages} {117205} (\bibinfo {year} {2013})}\BibitemShut {NoStop}%
\bibitem [{\citenamefont {Ciuciulkaite}\ \emph {et~al.}(2019)\citenamefont
  {Ciuciulkaite}, \citenamefont {\"Ostman}, \citenamefont {Brucas},
  \citenamefont {Kumar}, \citenamefont {Verschuuren}, \citenamefont
  {Svedlindh}, \citenamefont {Hj\"orvarsson},\ and\ \citenamefont
  {Kapaklis}}]{PhysRevB_Agne_FMR_2019}%
  \BibitemOpen
  \bibfield  {author} {\bibinfo {author} {\bibfnamefont {A.}~\bibnamefont
  {Ciuciulkaite}}, \bibinfo {author} {\bibfnamefont {E.}~\bibnamefont
  {\"Ostman}}, \bibinfo {author} {\bibfnamefont {R.}~\bibnamefont {Brucas}},
  \bibinfo {author} {\bibfnamefont {A.}~\bibnamefont {Kumar}}, \bibinfo
  {author} {\bibfnamefont {M.~A.}\ \bibnamefont {Verschuuren}}, \bibinfo
  {author} {\bibfnamefont {P.}~\bibnamefont {Svedlindh}}, \bibinfo {author}
  {\bibfnamefont {B.}~\bibnamefont {Hj\"orvarsson}}, \ and\ \bibinfo {author}
  {\bibfnamefont {V.}~\bibnamefont {Kapaklis}},\ }\href {\doibase
  10.1103/PhysRevB.99.184415} {\bibfield  {journal} {\bibinfo  {journal} {Phys.
  Rev. B}\ }\textbf {\bibinfo {volume} {99}},\ \bibinfo {pages} {184415}
  (\bibinfo {year} {2019})}\BibitemShut {NoStop}%
\bibitem [{\citenamefont {Pohlit}\ \emph {et~al.}(2019)\citenamefont {Pohlit},
  \citenamefont {Muscas}, \citenamefont {Chioar}, \citenamefont {Stopfel},
  \citenamefont {Ciuciulkaite}, \citenamefont {\"Ostman}, \citenamefont
  {Pappas}, \citenamefont {Stein}, \citenamefont {Hj\"orvarsson}, \citenamefont
  {J\"onsson},\ and\ \citenamefont {Kapaklis}}]{zenodo_susceptibility}%
  \BibitemOpen
  \bibfield  {author} {\bibinfo {author} {\bibfnamefont {M.}~\bibnamefont
  {Pohlit}}, \bibinfo {author} {\bibfnamefont {G.}~\bibnamefont {Muscas}},
  \bibinfo {author} {\bibfnamefont {I.-A.}\ \bibnamefont {Chioar}}, \bibinfo
  {author} {\bibfnamefont {H.}~\bibnamefont {Stopfel}}, \bibinfo {author}
  {\bibfnamefont {A.}~\bibnamefont {Ciuciulkaite}}, \bibinfo {author}
  {\bibfnamefont {E.}~\bibnamefont {\"Ostman}}, \bibinfo {author}
  {\bibfnamefont {S.}~\bibnamefont {Pappas}}, \bibinfo {author} {\bibfnamefont
  {A.}~\bibnamefont {Stein}}, \bibinfo {author} {\bibfnamefont
  {B.}~\bibnamefont {Hj\"orvarsson}}, \bibinfo {author} {\bibfnamefont {P.~E.}\
  \bibnamefont {J\"onsson}}, \ and\ \bibinfo {author} {\bibfnamefont
  {V.}~\bibnamefont {Kapaklis}},\ }\href {\doibase 10.5281/zenodo.3607135}
  {\enquote {\bibinfo {title} {Collective magnetic dynamics in artificial
  spin-ice probed by \relax{AC} susceptibility},}\ } (\bibinfo {year} {2019}),\
  \bibinfo {note} {\relax{Z}enodo,
  \url{https://doi.org/10.5281/zenodo.3607135}}\BibitemShut {NoStop}%
\end{thebibliography}

%

\end{document}